\documentclass[twocolumn,prl,showpacs,superscriptaddress,preprintnumbers,amsmath,amssymb]{revtex4}

\usepackage{graphicx}
\usepackage{dcolumn}
\usepackage{bm}
\usepackage[dvips]{color}
\usepackage{ulem}
\begin{document}
\title{Non-resonant inelastic x-ray scattering involving excitonic excitations}

\author{M. W. Haverkort}
  \affiliation{II. Physikalisches Institut, Universit{\"a}t zu K{\"o}ln, Z{\"u}lpicher Str. 77, D-50937 K{\"o}ln, Germany}
\author{A. Tanaka}
  \affiliation{Department of Quantum Matter, ADSM, Hiroshima University, Higashi-Hiroshima 739-8530, Japan}
\author{L. H. Tjeng}
  \affiliation{II. Physikalisches Institut, Universit{\"a}t zu K{\"o}ln, Z{\"u}lpicher Str. 77, D-50937 K{\"o}ln, Germany}
\author{G. A. Sawatzky}
  \affiliation{Department of Physics and Astronomy, University of British Columbia, Vancouver, British Columbia, Canada V6T 1Z1}

\date{\today}

\begin{abstract}

In a recent publication Larson \textit{et al.} \cite{Larson07} reported remarkably clear $d$-$d$ excitations for NiO and CoO measured with x-ray energies well below the transition metal $K$ edge. In this letter we demonstrate that we can obtain an accurate quantitative description based on a local many body approach. We find that the magnitude of $\vec{q}$ can be tuned for maximum  sensitivity for dipole, quadrupole, etc. excitations. We also find that the direction of $\vec{q}$ with respect to the crystal axes can be used as an equivalent to polarization similar to electron energy loss  spectroscopy, allowing for a determination of the local symmetry of the initial and final state based on selection rules. This method is more generally applicable and combined with the high resolution available, could be a powerful tool for the study of local distortions and symmetries in transition metal compounds including also buried interfaces.
\end{abstract}


\maketitle

Transition metal compounds with partially filled $d$-shells show a large variety of interesting properties, like metal insulator transitions, colossal magneto resistance and super conductivity \cite{Tsuda00, Imada98}. One reason for the complex behavior of these materials is the strong interplay between the orbital, charge, spin, and lattice  degrees of freedom \cite{OrbFocus04}. In the manganates, for example, the orbital and charge ordering could be related to the magneto resistance \cite{Tokura00}. For vanadium oxides it has been shown that the orbital occupation changes drastically at the metal insulator transition \cite{Park00,Haverkort05} and in the layered high temperature superconducting cuprates the two dimensional electronic structure is intimately linked to the unoccupied $d_{x^2-y^2}$ orbital. More recently there is a lot of activity in using interface induced effects in transition metal oxides to strongly modify the physical properties \cite{Ohtomo04, Okamoto04}.

Since the orbital degrees of freedom play an important role in all of these materials it is highly desirable to have good experimental methods to determine the energy scale of the crystal or ligand field splitting and the local symmetry. In principle this can be done by optical spectroscopy \cite{Newman59}, however these so called $d$-$d$ excitations are even and therefore optically forbidden and often completely masked by transitions involving small amounts of impurities or defects. The  reason why some of these transitions are optically visible at all is due to simultaneous excitations of magnons or phonons.  This results in an intensity typically 1000 times smaller than the intensity found for the close-by charge-transfer or Mott-Hubbard excitations. In multilayers and interfaces the problem is even more severe since there may be a variety of optical transitions due to other components which quickly mask out the $d$-$d$ transitions.  One is even  not always able to easily discriminate between absorption peaks due to $d$-$d$ excitations also referred to as orbiton excitations  and multiple phonon excitations for example \cite{Saitoh01, Grueninger02}.

Recently resonant inelastic x-ray scattering (RIXS) techniques have been developed to study $d$-$d$ excitations. At the transition metal $K$-edge a $1s$ to $4p$ excitation is involved. In the intermediate state the $d$ levels shift due to the changed local potential. This energy shift can change the occupied $d$ orbital wave function,  which may result in a resonant enhancement of the $d$-$d$ excitations \cite{Kotani01, Brink05, Platzman98}. The draw-back of resonant scattering at the $K$-edge is however, that charge-transfer excitations are enhanced much more efficiently than the $d$-$d$ \cite{Kao96, Kotani01}, for two reasons. The first is that the $4p$-orbitals of the intermediate state are quite spatially extended and have a small interaction with the $3d$ orbitals but a very large one with the surrounding O $2p$ orbitals. The second reason is that the {\it spherical} core hole potential does not enhance $d$-$d$ transitions directly. An other option developed recently is RIXS at the transition metal $L_{2,3}$ edge or $M_{2,3}$ edge. Here one excites (and de-excites) a $2p$ or $3p$ transition metal core electron into the $3d$ valence shell. With this technique one can choose which of the low-lying energy states one wants to enhance by selecting the incident energy and polarization \cite{Ghiringhelli05, Magnuson02, Kuiper98, Kotani01}.

A recent paper by Larson \textit{et al.} \cite{Larson07} exhibited clear $d$-$d$ excitations within the gap in NiO and CoO with the use of non-resonant inelastic x-ray scattering (NIXS) for energies just below the $K$ edge. In principle one should expect that these excitations can be seen with NIXS, but surprisingly they found that the intensity of the $d$-$d$ excitations at certain $\vec{q}$ vectors is much higher than the intensity of the charge-transfer or Mott-Hubbard excitations. They analyzed their experimental findings in the framework of LDA+U, which describes the transitions in terms of a one particle interband transition rather than a transition involving strongly bound excitonic states as is known to be the case for these states in NiO and CoO. Very interesting to note though is that the angular dependent results can be quite well described within the band structure approach for the case of NiO because the transition involves basically a promotion of a $t_{2g}$ electron into an unoccupied $e_{g}$ state. In more complicated cases involving multi Slater determinant excitonic bound states the situation will be quite a bit more involved as also realized by the authors of that paper and previous work on the cuprates by Ku \textit{et al.} \cite{Ku02}.

In this letter we will develop a local but many body treatment of NIXS and describe the observed $d$-$d$ excitations within a configuration interaction cluster calculation analogous to the approaches used for analyzing the energy positions of optical $d$-$d$ excitations \cite{Janssen88} or the RIXS spectra at the $L_{2,3}$ edge \cite{Ghiringhelli05, Magnuson02}. The goal is to show that we can have a straightforward and quantitative description of the NIXS process and that this will open up new opportunities to extract detailed and invaluable information concerning the local electronic structure of correlated electron systems not easily accessible by other techniques.

The interaction of matter with light is given by two terms. One proportional to the vector potential ($\vec{A}$) squared, the other proportional to the dot product of the momentum operator for the electrons ($\vec{p}$) with the vector potential.
\begin{equation}
H_{int}=\frac{e^2}{2 m_{e} c^2} \vec{A}^2 + \frac{e}{m_e c} \vec{p} \cdot \vec{A}
\end{equation}
At resonance the second term ($\vec{p}\cdot \vec{A})$ is responsible for the largest contribution to the scattering cross section. Off resonance however this term looses importance rapidly  and the scattering is governed mainly by the term $\vec{A}^2$. The off resonance scattering cross section is then given by
\begin{eqnarray}
\nonumber  \frac{d^2\sigma}{d\Omega d\omega_f} &=& r_0^2 \frac{\omega_{f}}{\omega_{i}} \sum_f \left| \vec{\epsilon}_i.\vec{\epsilon}_f^*  \langle f| e^{\imath(\overrightarrow{k_i}-\overrightarrow{k_f})\cdot \overrightarrow{r}} |i \rangle \right|^2\\
            &&  \delta(E_{i}-E_{f}+\hbar(\omega_i-\omega_f))
\end{eqnarray}
One can define the dynamical structure factor $S(\vec{q},\omega)$, which is a function of the scattering vector $\vec{q}=\vec{k}_i-\vec{k}_f$ and the energy loss $\omega=\omega_i-\omega_f$ as $S(\vec{q},\omega)=\frac{d^2\sigma}{d\Omega d\omega_f} / (r_0^2 \frac{\omega_{f}}{\omega_{i}} \left| \vec{\epsilon}_i.\vec{\epsilon}_f^* \right|^2)$, which has the advantage that all non-material dependent factors are factored out.

$S(\vec{q},\omega)$ is a sum over transition probabilities multiplied by a delta function responsible for the energy conservation. This can be written as a Greens function in the spectral representation:
\begin{eqnarray}
\nonumber  S(\vec{q},\omega)&=&\sum_f \left| \langle f| e^{\imath \overrightarrow{q}\cdot \overrightarrow{r}} |i \rangle \right|^2 \delta(E_{i}-E_{f}+\hbar \omega)\\
&=&\lim_{\Gamma \rightarrow 0} -\frac{1}{\pi} \textrm{Im} \langle i | T^{\dag} \frac{1}{E_{i}-H+\hbar\omega+\frac{\imath \Gamma}{2}} T|i\rangle
\end{eqnarray}
With the transition matrix equal to $T=e^{\imath \overrightarrow{q}\cdot \overrightarrow{r}}$.

To enable a rather direct symmetry analysis we prefer to  discuss the transitions in terms of monopole, dipole, quadrupole, etc. excitations and in order to do so we expand the transition matrix on spherical harmonics.
\begin{eqnarray}
\nonumber  T&=&e^{\imath \overrightarrow{q}\cdot \overrightarrow{r}}\\
\nonumber  &=&\sum_{k=0}^{\infty} \sum_{m=-k}^{k}  \imath^k (2k+1) j_{k}(q\,r)\\
&& \qquad\qquad\times{C_{m}^{(k)}}^*(\theta_{q},\phi_{q}) C_{m}^{(k)}(\theta_{r},\phi_{r})
\end{eqnarray}
with $C_m^{(k)}=\sqrt{\frac{4 \pi}{2k+1}}Y_{km}$, and $Y_{k,m}$ the spherical harmonics. This results in a sum over $k$ of a spherical Bessel function of order $k$ times a spherical harmonic of order $k$. For $d$-$d$ excitations only monopole ($k$=0), quadrupole ($k$=2) and hexadecimalpole ($k$=4) transitions are allowed and therefore only three values of $k$ have to be evaluated. We first discuss the effect of changing the length of the $\vec{q}$ vector, which enters via the expectation value of the spherical Bessel function over $q \, r$ and second discuss the effect of changing the orientation of the sample with respect to the $\vec{q}$ vector which enters as a spherical harmonic over the angular coordinates of $\vec{q}$.

\begin{figure}
    \includegraphics[width=0.45\textwidth]{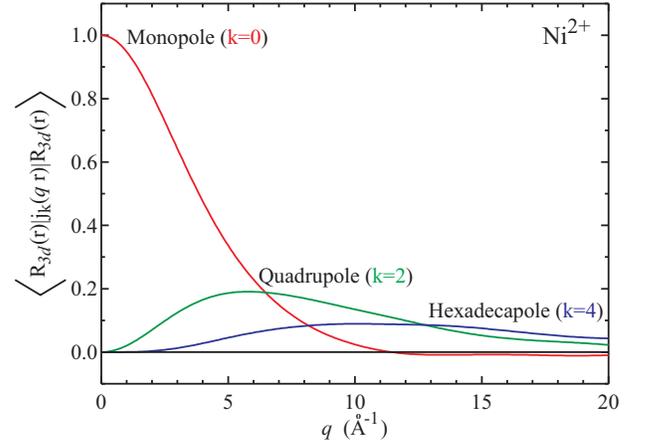}
    \caption{(color online) Expectation value of the spherical Bessel function for monopole, quadrupole and hexadecapole transitions as a function of $q$.}
    \label{fig1}
\end{figure}

The expectation value of a spherical Bessel function becomes small if the spherical Bessel function oscillates many times over the length scale of the product of the initial and final state wave function. We therefore expect the maximum intensity to occur at $q$ values corresponding to the period of a spherical Bessel function of length comparable to the atomic radial extent of the $d$ wave function. In figure 1 we plot the expectation value of the spherical Bessel function for $k$=0, 2 and 4 calculated for Ni$^{2+}$ where the radial wave function has been calculated within the Hartree-Fock approximation with the use of Cowan's code \cite{Cowan81}. One can clearly see that each of the different multipoles has a maximum at a different $\vec{q}$ vector. For monopole excitations one should use small wave vectors or forward scattering although a monopole transition contributes only to the zero energy loss peak because the excited states are orthogonal to the ground state since they are eigenfunctions of the same Hamiltonian. Quadrupole excitations become maximal around 5 {\AA}$^{-1}$ and hexadecapole excitations become largest between 8 and 14 {\AA}$^{-1}$. This is great as it allows one to choose which excitation one wants to measure.

\begin{figure}
    \includegraphics[width=0.45\textwidth]{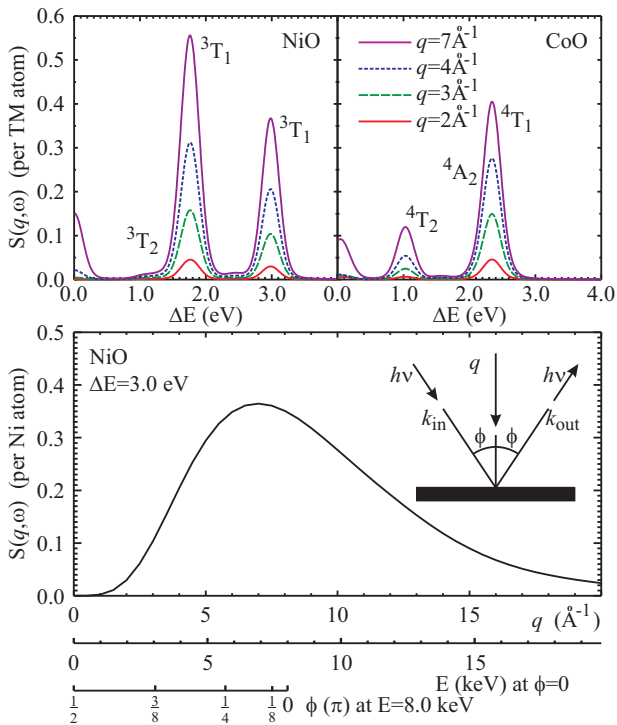}
    \caption{(color online) Top panels; NIXS spectra for different values of $q$ in the [111] direction calculated for a NiO$_{6}$ and CoO$_{6}$ cluster. The monopole scattering has not been included. Bottom panel; NIXS intensity for the NiO peak at 3.0 eV loss calculated for different values of $q$. The photon energy at $\phi$=0 and the scattering angle ($\phi$) at a photon energy of 8 keV are given as alternative scales.}
    \label{fig2}
\end{figure}

In the top panels of figure 2 we show the NIXS spectra of NiO and CoO calculated for a TMO$_{6}^{10-}$ cluster consisting of a transition-metal ion and surrounding six oxygen ions, with the use of the program XTLS8.3 \cite{Tanaka94, parrameters}. For both NiO and CoO we see two peaks, with maximum intensity for $q$=7{\AA}$^{-1}$ in good agrement with the measurements of Larson \textit{et al.} \cite{Larson07}. The $d$-$d$ excitations are labeled by the symmetry of the final-state without the inclusion of spin-orbit coupling. These peaks are split by spin-orbit coupling, as states of $T$ symmetry are 3-fold orbital degenerate. Spin-orbit coupling has been included for the calculations, but the splitting can not be resolved with this resolution. Better experimental resolutions is possible and it would be interesting to look at these excitations with higher resolution. We now can compare these calculations to the $d$-$d$ spectra found in optical spectroscopy \cite{Newman59} or RIXS \cite{Ghiringhelli05} at the $L_{2,3}$ edge. The first thing one notices is that with optics and RIXS one sees many more $d$-$d$ excitations than with NIXS. The explanation is straightforward  if one considers the selection rules. For NIXS one has pure charge excitations and therefore the selection rule $\Delta S=0$. For NiO which has a ground-state of $^{3}A_{2}$ symmetry with $t_{2g}^6 e_{g}^2$ configuration \cite{Ballhausen62} there are three possible excited states that are also triplets, namely two states of $^{3}T_{1}$ symmetry (around 1.8 and 3.0 eV) and one of $^{3}T_{2}$ symmetry (around 1.1 eV). One can see however only two peaks as the $^{3}T_{2}$ state can not be reached with a quadrupole excitation. These selection rules are rather different in RIXS at the $L_{2,3}$ edge. There one has an intermediate state with a core hole in the $2p$ shell of the transition metal. The spin-orbit coupling constant for $2p$-core electrons of Ni is around 11.5 eV and mixes states of different spin. This mixing results in different spin state transitions to be observed with comparable intensities in RIXS at the $2p$ edge.

In the bottom panel of figure 2 we show the NIXS intensity of the 3.0 eV loss peak as a function of the magnitude of $q$. There are several ways in which one could change the magnitude of $q$. For the geometry as shown in the inset of figure 2, $|\vec{q}|=2\,\cos(\phi)\frac{2 \pi E}{h c}$. Which means one can change the energy of the photons or change the scattering angle in order to change the magnitude of $q$. For convenience we show three different, equivalent scales for the bottom panel of figure 2.

\begin{figure}
    \includegraphics[width=0.45\textwidth]{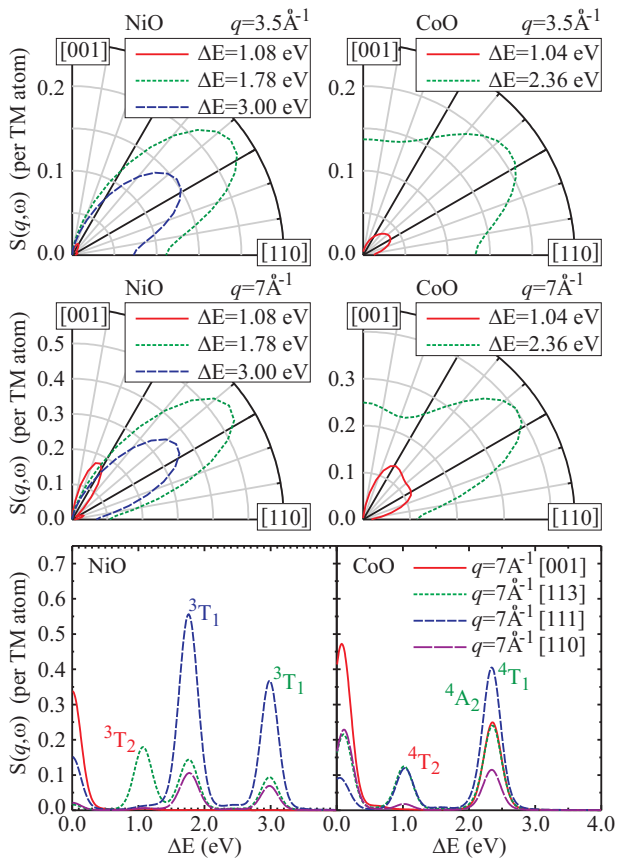}
    \caption{(color online) Top panels; Angular dependence of the NIXS intensity for NiO and CoO at different loss energies calculated at $q$=3.5 {\AA}$^{-1}$ and at $q$=7 {\AA}$^{-1}$. Bottom panels; NIXS spectra at $q$=7 {\AA}$^{-1}$ for different sample orientations.}
    \label{fig3}
\end{figure}

Another advantage of NIXS is that one can not only tune the magnitude of $q$ in order to optimize the scattered intensity one can also use the directional dependence, i.e. the direction of the $\vec{q}$ vector with respect to the crystal axes in order to do something equivalent to polarization analyzes. The transition matrix depends on the direction of the $\vec{q}$ vector by, $\sum_{m=-k}^{k} {C_{m}^{(k)}}^*(\theta_{q},\phi_{q}) C_{m}^{(k)}(\theta_{r},\phi_{r})$. For a dipole transition ($k=1$) for example this is equivalent to a dipole in the direction of $\vec{q}$. This allows for a determination of the symmetry of the initial and the final state, based on selection rules. In the top panels of figure 3 we show the angular dependence for the NIXS intensity of different energy loss peaks of NiO and CoO at $q$=3.5 {\AA}$^{-1}$ and $q$=7 {\AA}$^{-1}$. One can see that the two peaks of $^3T_1$ symmetry in NiO show the same angular dependence whereas the two peaks in CoO, which are of different symmetry show a different angular dependence. It should be noted that the angular dependence calculated at $q$=3.5 {\AA}$^{-1}$ for the peaks at 3.0 (2.36) eV energy loss of NiO (CoO) show good agrement with the intensities as measured by Larson \textit{et al.} \cite{Larson07}. It is interesting to note that the $d$-$d$ excitation at 1.1 eV in NiO, which is not quadrupole allowed can be seen at $q$=7 {\AA}$^{-1}$, with the use of a hexadecapole transition. These are strongly peaked in approximately the $[113]$ direction.

To conclude we have expanded the non-resonant contribution ($A^2$) to the dynamical structure factor ($S(\vec{q},\omega)$) in spherical harmonics. $S(\vec{q},\omega)$ for CoO and NiO has been calculated with the use of this expansion. We used a configuration interaction cluster calculation for a NiO$_{6}^{10-}$ and CoO$_{6}^{10-}$ cluster, in order to describe the final-state excitons correctly. The calculated spectra are in excellent agreement with measurements of Larson \textit{et al.} \cite{Larson07}. The spectral representation of $S(\vec{q},\omega)$ presented here gives a straightforward explanation of the measured energy loss intensity. A big advantage is that $S(\vec{q},\omega)$ in the multipole expansion is easy to calculate. This is especially suitable for $q$ values comparable to atomic dimensions. For larger energy transfers involving interband transitions and collective modes one could rely on LDA+U or time dependent DFT \cite{Ku02, Eguiluz05}. By changing the magnitude of $q$, one can tune the sensitivity of the measurement to different multipoles and optimize the intensity of the $d$-$d$ excitation. A certain multipole has optimal intensity if the spherical Bessel function of the same order has a period comparable to the size of the local $d$ orbital. Rotating the sample with respect to the $\vec{q}$ vector allows one to do something equivalent to polarization analysis in \textit{normal} spectroscopy. This creates the opportunity to determine the symmetry of the ground-state and excited-state with the use of selection rules. It is important to note that this kind of measurement is bulk sensitive and can be used to study buried interfaces. The elemental sensitivity is not as strong as in RIXS but because the radial matrix elements depend strongly on the radial extend of the $d$ wave functions some degree of elemental sensitivity remains. We believe that this kind of measurement can provide important information on the electronic structure and local symmetry of some of the most fascinated strongly correlated electron systems.

We would like to thank Wei Ku and B. C. Larson for helpful discussions and for the use of their data. Investigation of the fine structure of the spectra was partly motivated by unpublished work by Cai \textit{et al.} and Baron \textit{et al.} This work was supported by the Deutsche Forschungsgemeinschaft through SFB 608 and the Canadian funding agencies NSERC, CIAR, and CFI.


\begin{thebibliography}{99}
\bibitem{parrameters} The parameters used for the cluster calculations are equivalent to the ones used for the RIXS calculations \cite{Ghiringhelli05}.
\end{thebibliography}
\end{document}